# Inductively Coupled Circuits with Spin Wave Bus for Information Processing

[1)]A. Khitun, [1)] M. Bao, [1)]J-Y. Lee, [1)]K. L. Wang

[2)] D.W. Lee, [2)] S. Wang, and [3)] Igor V. Roshchin

[1)] Device Research Laboratory, Electrical Engineering Department,

MARCO Focus Center on Functional Engineered Nano Architectonics (FENA),

University of California at Los Angeles, Los Angeles, California, 90095-1594

[2)] Stanford University, Geballe Laboratory for Advanced Materials, Department of Materials Science and Engineering, Stanford University, Stanford, California, 94305-4045

[3)] Physics Department, University of California, San Diego, La Jolla, CA 92093-0319

## Abstract

We describe a new approach to logic devices interconnection by the inductive coupling via a ferromagnetic film. The information among the distant devices is transmitted in a wireless manner via a magnetic field produced by spin waves propagating in the ferromagnetic film, referred to as the spin wave bus. As an alternative approach to the transistor-based architecture, logic circuits with spin wave bus do not use charge as an information carrier. A bit of information can be encoded into the phase or the amplitude of the spin wave signal. We present experimental data demonstrating inductive coupling through the 100nm thick NiFe and CoFe films at room temperature. The performance of logic circuits with spin wave bus is illustrated by numerical modeling based on the experimental data. Potentially, logic circuits with spin wave bus may resolve the interconnect problem and provide "wireless" read-in and read-out. Another expected benefit is in the enhanced logic functionality. A set of “NOT”, “AND”, and “OR” logic gates can be realized in one device structure. The coupling between the circuits and the spin wave bus can be controlled. We present the results of numerical simulations showing the controllable switching of a bi-stable logic cell inductively coupled to the spin wave bus. The shortcomings and limitations of circuits with spin wave bus are also discussed.

## I. Introduction:

There is great interest in novel logic circuits [1] that are aimed to provide high information/signal processing rates and that are scalable to the nanometer range. The development of novel functional materials and structures at nanometer scale, together with processes for monolithic, heterogeneous integration with CMOS, may result in a breakthrough for revolutionary new types of electronic circuitry with capabilities far beyond the traditional CMOS platform. Today, one of the major obstacles for nano-circuitry is the interconnect problem. The growing number of devices per unit area results in tremendous difficulties with interconnection wiring [2]. Impedance match between devices and wires is another serious issue. It would be of great benefits to construct scalable logic circuits with wireless communication among the nanoscale devices. In this work, we present the concept of inductively coupled circuits, where the information exchange among the elementary functional units is accomplished in a wireless manner via inductive voltage signals produced by spin waves propagating in a ferromagnetic film.

Spin wave as a physical phenomenon has attracted scientific interest for a long time [3]. Spin wave is a collective oscillation of spins in an ordered spin lattice around the direction of magnetization. The phenomenon is similar to the lattice vibration, where atoms oscillate around their equilibrium position. Potentially, it is possible to use ferromagnetic films as a conduit for spin wave propagation, referred to as "Spin Wave Bus" (SWB), where the information can be coded into a phase or the amplitude of the spin wave. The distinct feature and key advantage of the SWB is that information transmission is accomplished without electron transport. Besides, there are other significant advantages: (i) ability to use superposition of spin waves in the bus to achieve useful logic functionality; (ii) a number of spin waves with different frequencies can be simultaneously transmitted among a number of spin-based devices; (iii) the interaction between spin waves and outer devices can be done in a wireless manner, via a magnetic field. The excitation of spin waves can be done by the local magnetic field produced by micro- or nano-scale antenna, while the detection of the spin wave is via the inductive voltage produced by propagating spin waves. The time-resolved inductive voltage measurement technique has been successfully applied for spin wave propagation study in nanoscale ferromagnetic films [4, 5]. It was demonstrated that spin waves in NiFe films can transmit a millivolt inductive voltage signal for 50μm distance at room temperature [5]. Spin waves can be also implemented in different spin-based nano architectures [6]. A first working spin-wave based logic circuit has been recently experimentally demonstrated [7]. The aim of this work is to present experimental data on inductive coupling through ferromagnetic films and to demonstrate the potential advantages of using inductively coupled circuits with SWB for information processing.

The paper is organized as follows: in the next section we explain the general concept of inductively coupled circuits with SWB. In Section III, we describe the experimental setup and present experimental data showing the inductive coupling through a 100 nm thick ferromagnetic films at room temperature. In Section IV, we present the results of numerical modeling illustrating the performance of logic circuits with SWB. Discussion and Conclusions are given in sections V and VI, respectively.

## II. Principle of Operation

The general idea of using inductively coupled circuits for information processing is to transmit information between the circuits using a magnetic flux without transmitting current via wires. According to the Faraday's law, the change of the magnetic flux through the surface generates an inductive voltage proportional to the rate of the magnetic flux change $E_{ind} = -d\Phi_m / dt$. In order to direct the magnetic flux among the circuits, we propose to use magnetic materials (ferromagnetic, anti-ferromagnetic, or ferrite). The change of magnetization in the magnetic material alters the outgoing magnetic flux and results in the inductive voltage signal. There are two possible ways to change magnetization through a magnetic material: (i) by domain wall motion [8] or (ii) by propagation of a spin wave without changing the domain orientation [6]. In both cases the magnetization change is not accompanied by charge transmission. We consider spin wave as an energetically efficient approach as it takes much less energy to excite a spin wave than to reverse the magnetization of the entire domain. Spin wave is confined within a magnetic film and can be efficiently guided. The film serves as a "magnetic" waveguide transmitting information coded in a spin wave from one circuit to another.

The material structure of a prototype logic circuit with SWB is shown in Fig.1(a). The core of the structure consists of a ferromagnetic film (NiFe, for example) deposited on a semi-insulating substrate, e.g. silicon on insulator (SOI) by a sputtering technique. The film may be entirely magnetized along the X axis when a magnetic field is applied. The thickness of the ferromagnetic layer is about 10 nm to 100 nm. There are three **a**symmetric **c**o**p**lanar **s**trip (ACPS) transmission lines on the top of the structure. These transmission lines are isolated from the ferromagnetic layer by the silicon oxide layer, and are used as input/output ports. A voltage pulse applied to an input ACPS line produces a magnetic field perpendicular to the magnetization of the ferromagnetic film, and, thus, generates spin waves. The readout can be done by an output ACPS line. The structure is similar to the one used for time-resolved measurements of propagating spin waves [5]. Fig.1(b) displays the equivalent circuit for the proposed logic circuit. The input and output circuits consisting of the spin-based devices shown in (a) are depicted as LCR oscillators $(L_1, C_1, R_1)$. The LCR transmission line $(L_0, C_0, R_0)$ is represented as the SWB. The oscillators are inductively coupled via the interaction with the ferromagnetic film. The change of the current in any of the oscillators produces an inductive voltage in the others, and vice versa. Depending on the relative phase of the currents in the input circuits, the output inductive voltage at the central circuit can be maximal (in phase) or minimal (out of phase).

As we will show, an elementary logic gates such as "NOT", "AND" and "OR" can be realized as from the prototype logic circuit shown in Fig.1(a). For example, the edge ACPS lines can be considered as the input ports, and the middle ACPS as the output port. The middle ACPS line detects the inductive voltage produced by the *superposition* of two waves. The input data are received in the form of voltage pulses. For simplicity, we assume that the input signal amplitudes of +1V and –1V correspond to the logic states 1 and 0, respectively. Next, the input information is encoded into the phase of the spin wave. The conversion of the voltage signal into the spin wave phase is accomplished by the microstrip. Each microstrip generates a local magnetic field to excite a spin wave in a ferromagnetic film. Depending on the polarity of the input signal, the initial phase of each spin wave may have a relative phase difference of e.g. $\pi$. Phases of "0" and "$\pi$" may be used to represent two logic states 1 and 0. Depending on the relative phase of the spin waves, the amplitude of the inductive voltage may be enhanced (in phase) or decreased (out of phase) compared to the inductive voltage produced by a single spin wave. Then, the voltage signal may be amplified by conventional MOSFET to provide the compatibility with the external circuits. The relative phase is defined by the location of the ACPS lines and the polarity of the input excitation voltage signal. It is also possible to control the initial phase by adjusting the time of excitation. With these elementary sets of devices and bus, it is possible to realize different logic gates AND, OR, and NOT by controlling the *relative phases* of the spin waves.

## III. Experimental data

In order to investigate the coupling via ferromagnetic films, we started with a test structure shown in Fig.2(a). There are two microstrips on the top of the structure. One microstrip is the transducer to excite spin waves, and the other one is the receiver to detect the inductive voltage. In our experiments, we used CoFe and NiFe ferromagnetic films. The ferromagnetic films have been fabricated as follows: A film of $Ni_{80}Fe_{20}$ (99.99%) was deposited at a rate of 0.05 nm/s using an electron-beam evaporator. The thickness of all the films was determined to be 95±4 nm with small-angle X-ray reflectometry. The films are polycrystalline with no magnetic easy axis. SQUID measurements indicate a small (~10 Oe or smaller) coercivity at room temperature.

A $Co_{30}Fe_{70}$ (atomic %) film was deposited using a Perkin-Elmer high vacuum rf-sputtering system and the film exhibited a saturation magnetization ($B_s$) of ~2.2 T. While a static magnetic field of 50 Oe was applied during the deposition to align the easy axis, the B-H measurement after the deposition gave an in-plane isotropic square-like hysteresis loop with a coercivity ($H_c$) of ~101 Oe. In order to improve the magnetic alignment, a field annealing was applied after the deposition at 270C for 10 hours with an applied magnetic field of 1 T along the easy axis in vacuum ($\sim 1\cdot 10^{-6}$ Torr). The $H_c$ values were decreased to 23 Oe and 11 Oe along the easy and hard axes, respectively, after the field annealing.

The structure was patterned and etched by reactive ion etch (RIE) to form a mesa structure. After the mesa formation and another 450 nm CVD-$SiO_2$ deposition on top of the patterned samples, metal transmission lines (500 nm-thick gold wires) were formed with various sizes of line width and spacing. The distance between the microstrips is 4μm. The experimental setup is shown in Fig.2(b). We use a pulse generator (Picosecond 4000B) to apply a voltage signal to the excitation microstrip. The inductive voltage in the receiving microstrip was detected by a high-frequency 500 MHz oscilloscope. The impedance of the excitation microstrip is matched to that of the outer circuit (50 Ω), to provide the maximum current.

In our first experiment we obtained the time-domain data on spin wave propagation in 100nm thick CoFe film. The excitation voltage pulse has the rise time of 100 ps and the pulse voltage of 2.5 V. In order to eliminate the direct electromagnetic coupling and obtain a pure spin-wave response, the two measured spectra of with and without magnetic filed are subtracted from each other. In Fig.3 we present the experimental data for the sample with CoFe film obtained after the subtraction (signal obtained for $H_{ext}$=0 is subtracted from $H_{ext}$=50Oe). In Fig.4 we show results for a similar experiment with a NiFe film obtained after the subtraction (signal obtained for $H_{ext}$=0 is subtracted from $H_{ext}$=300Oe). These graphs illustrate the spin wave signal as a result of voltage pulse excitation. In both cases, the inductive voltage has a form of high frequency (ω=2.5GHz) damping voltage oscillations, with a characteristics decay time $\tau \sim$ 0.6ns. The results for spin wave extraction are consistent with the previously reported one in Ref. [4, 5] for permalloy films. All measurements were performed at room temperature.

Next, we carried out experiments to investigate the coupling when the input signal is a mix of two frequencies, $f_1$ and $f_2$. In Fig.5 we show the experimental setup. Two continuous signals at 6.2GHz and 6.9GHz were generated by Agilent 83711B and Agilent 83640B signal generators. A set of RF devices were used to mix and to amplify the combined signal. Two band pass filters BF1 and BF2 (6.16-7.0 GHz window) are used to cutoff mixed frequencies produced by TWT power amplifier. The output signal was detected by the Agilent 8592A spectrum analyzer. In Fig.6 we plot the power of the detected signal at different frequencies. The frequency conversion was observed, where the output spectrum contains the mixed frequencies of the two inputs. In Fig.6(a), the output signal spectrum in the range of 3-15GHz is shown. The two most prominent frequency responses (-11dBm) observed correspond to the excitation frequencies 6.2

GHz and 6.9 GHz. Signal with a lower power was observed at the frequencies that correspond to the second harmonics of the input signal and the linear combinations of the input signal and the second harmonics, $2f_2 - f_1$, $2f_1$, $f_1 + f_2$, $2f_2$, *and* $2f_1 - f_2$. A much lower power signal (-62 dBm) detected at the frequency $f_2 - f_1$ is presented in Fig.6(b). We attribute this frequency conversion to the non-linear magnetization response of the ferromagnetic material to the external excitation magnetic field produced by the microstrip. The appearance of the mixed frequencies gives us an additional possibility to achieve useful logic functionality for the proposed inductively coupled circuits.

## IV. Logic Performance Simulations

Information to be transmitted among the inductively coupled circuits can be encoded in two possible ways. First, one can use the phase of the inductive voltage signal. For example, two relative phases 0 and $\pi$ can be used to represent the logic states 0 and 1, respectively. In this case, all coupled circuits operate at the same frequency and computations are performed by simply the superposition of different input waves. In an earlier work, we presented numerical simulations showing the inductive voltage produced by the superposition of two spin wave packets [6]. The inductive voltage is maximum if two waves are in phase (twice the amplitude of the local magnetization change), and the inductive voltage is minimum if the two waves are out of phase and cancel each other. Second, we can represent logic states 0 and 1 by using the amplitude (power) of the inductive voltage signal at a particular frequency. In this case, different bits can be assigned to different frequencies. Below, we present the results of numerical simulations illustrating logic performance.

The general approach to the inductive coupling modeling in the proposed circuit structure includes three major components: (i) input circuit coupling to the spin wave bus, (ii) signal propagation through the ferromagnetic film, (iii) output circuit coupling to the spin wave bus. The coupling of the input circuit to the spin wave bus and the magnetization dynamics in the ferromagnetic film can be described by the Landau-Lifshitz-Gilbert equation as follows:

$$\frac{d\vec{m}}{dt} = -\frac{\gamma}{1+\alpha^2}\vec{m}\times\left[\vec{H}_{eff} + \alpha\vec{m}\times\vec{H}_{eff}\right], \qquad (1)$$

where $\vec{m} = \vec{M}/M_s$ is the unit magnetization vector, $M_s$ is the saturation magnetization, γ is the gyro-magnetic ratio, and $\alpha$ is the phenomenological Gilbert damping coefficient. The first term of equation (1) describes the precession of magnetization about an effective field ($\vec{H}_{eff}$) and the second term describes dissipation of this precession (or damping). The effective field, $\vec{H}_{eff}$, is given below:

$$\vec{H}_{eff} = \vec{H}_d + \frac{2A}{M_s}\nabla^2\vec{m} + \frac{2K}{M_s}(\vec{m}\cdot\vec{c})\vec{c} + \vec{H}_{ext}, \qquad (2)$$

where $\vec{H}_d = -\nabla\Phi$ and $\nabla^2\Phi = 4\pi M_s\nabla\cdot\vec{m}$; $A$ is the exchange constant, $K$ is the uniaxial anisotropy constant, $\vec{c}$ is the unit vector along with the uniaxial direction, and $\vec{H}_{ext}$ is the external magnetic field. Then, the inductive voltage produced in the output circuit can be found according to Ref.[4] as follows:

$$V_{ind} = \left( \frac{\mu_0 ldf(z,y)}{4} \right) \left( \frac{Z}{Z + 0.5R_{dc}} \right) \frac{d\overline{M}_y}{dt}, \tag{3}$$

where $\mu_0$ is the vacuum permeability, $l$ is the length of the ACPS line, $f(z,w)$ is the spatial loss function, $Z$ is the strip line impedance, and $R_{dc}$ is the total ACPS line dc resistance.

In principle, a rigorous simulation would require detailed analysis of the magnetic field produced by microstrip, calculation of the magnetization change produced in the ferromagnetic film as well as the propagation of the spin wave packet, and the inductive voltage calculations. The high-fidelity simulations are important for better understanding of the coupling phenomena and for circuit performance optimization. However, in order to illustrate the logic performance, we use a simplified approach and express the correlation between the input and output voltage signals in a convenient empirical form:

$$V_{ind} = \Gamma \cdot V_{in} \cdot e^{-t/\tau} \sin(\omega t + \varphi), \tag{4}$$

where $V_{in}$ is the peak input voltage applied to the transducer, $\Gamma$ is the coupling coefficient, $\tau$ is the attenuation time, $\omega$ is the frequency, and $\varphi$ is the initial phase of the inductive voltage signal. The parameters can be extracted from the time-domain experimental data given in figures 3 and 4.

As the amplitude of the output inductive voltage signal scales with the number of superposed spin waves, it provides a tool for *analog* computation. In order to make computation in a *digital* form, the circuit structure has to be modified by introduction of a non-linear element. In Fig.7(a) we depict a network, where each logic circuit consists of a pair of resonant tunnel diodes (RTDs) inductively coupled to the SWB. In our numerical simulations, we use a piece-wise approximation of the RTD *I-V* characteristic shown in Fig.7(b). There are three linear portions in the approximated curve, labeled as *b1*, *b2* and *b3*. $V_p$, $I_p$, $V_v$, $I_v$ are the peak voltage, the peak current, the valley voltage, and the valley current, respectively. The slopes of the branches *b1* and *b3* are assumed the same. The voltage and current are plotted normalized to $V_p$ and $I_p$, respectively. At a constant bias $V_{DD}$, each circuit may have two possible stable states: two output voltages "low"- $V/V_p$=0.5 and "high"- $V/V_p$=2.5, corresponding to logic states 0 and 1, respectively. The switching between the states can be achieved by the inductive voltage pulse. In the simple case, the switching occurs if the amplitude of the inductive voltage exceeds 2.0 $V_p$. We will show that the switching can be controlled by an external parameter - $V_{DD}$, making the circuits sensitive to the inductive voltage signal depending on the $V_{DD}$ signal.

In Fig.8 we present the results of numerical simulations showing the possibility of switching of the bi-stable RTD-based circuit by the inductive voltage signal. The switching is achieved by changing the bias voltage $V_{DD}$. In Fig. 8(a) the solid line corresponds to the output voltage. The dashed line labels the bias voltage $V_{DD}$. The input inductive voltage is shown in the inset. All voltages are normalized to $V_p$, and the time is normalized to RC ($R=V_p/I_p$ and C is the RTD's capacitance). For the case of constant $V_{DD}$ =3$V_p$, the amplitude of the inductive voltage is too small to switch the state of the RTD-based cell, as seen in the top graph. In Fig.8(b) we show the results of numerical simulations for the same circuit and the same inductive voltage signal when a time-dependent $V_{DD}$ (dashed line) is applied. The bias voltage is chosen to have one cycle of oscillation at the same frequency as the input inductive voltage signal. The bias oscillation and the inductive voltage signal are synchronized in time. In this case, the circuit switches its state from the "low" to the "high" output voltage. By adjusting the parameters of the RTD-based circuit with the parameters of the SWB, it is possible to realize a network where the switching of an individual bi-stable cell takes place if and only if there are both inductive voltage signal and bias voltage oscillations which are synchronized. Following this approach, it is possible to selectively couple a specific individual circuit to the SWB, making some of the circuit sensitive to the transmitted signal and some not at different periods of time.

There is a set of logic gates that can be realized using inductively coupled circuits in one device structure. In Fig.9 we show possible "NOT" and "AND" gates. The simplest inverter is a delay line, where the time of signal propagation in the SWB between the input and output ports is adjusted to provide π phase difference. In this case the sign of output signal is always opposite to the input one. The "AND" gate can be realized in two possible ways: using the effect of spin wave superposition or the frequency conversion. As the amplitude of the inductive voltage scales with the number of signals coming in phase (see Fig.7, the parameters of the output circuit can be adjusted so the switching takes place only at twice the signal amplitude. In other words, the amplitude of a single signal (or two signals coming out of phase) is not sufficient to switch the output state from "low" to "high". The switching occurs only if two signals are coming in phase. The "OR" gate can be realize by analogy with the "AND" gate by adjusting the circuit parameters and circuit coupling synchronization. Another possible realization of the "AND" gate is to use input ports tuned to different frequencies $f_1$ and $f_2$, respectively, and the output port tuned to frequency $f_1+f_2$. The output signal at frequency $f_1+f_2$ is nonzero if and only if there are nonzero inputs at both frequencies $f_1$ and $f_2$.

With "AND", "OR" and "NOT" gates, one can implement any logic function, and all these gates can be realized in one device structure.

## V. Discussion

The experimental data shows the prominent inductive coupling via ferromagnetic films at room temperature. At least several millivolts amplitude inductive voltage signal can be transmitted to a distance over micrometers in a wireless manner. The transmitted signal has a form of damping harmonic oscillations, whose phase correlates with the input voltage signal. Another interesting phenomenon is the frequency conversion observed for the mix input signal consisting of two frequencies. Based on these results, we have described a set of possible logic gates allowing us to realize any logic function for data processing.

There are many potential advantages for inductively coupled logic circuits with SWB for information processing. (i) The coupling between the circuits and the SWB can be done in a wireless manner, using micro- or nanometer scale strips. It may potentially resolve the interconnectivity problem faced today's IC's, as there are no conducting wires needed to unite the circuits. (ii) The use of spin waves for information transmission through ferromagnetic films eliminates the need in using electric current for this purpose which may be beneficial from the power consumption point of view. (iii) Transmitting signal in a waveform allow us to encode information in the phase of the propagating spin wave signal. Various logic gates ("NOT", "AND" and "OR") can be realized in one device structure by controlling the relative phase of spin wave signals. The utilization of wave superposition offers the possibility to accomplish useful information processing inside the bus, without the use of additional logic elements. (iv) The observed frequency conversion phenomenon provides an alternative approach to the logic gate construction. Potentially, it is possible to build a multi-bit processor, where a number of bits assigned to different frequencies can be processed in parallel. (v) The coupling between an elementary circuit and the SWB can be controlled (by an external parameter as shown in our numerical simulations). The combination of the RTD-based logic cells with SWB preserves all advantages of the Cellular Non-linear Network for image processing [9]. (vi) There is no major departure from CMOS fabrication. The inductively coupled circuits can be built on silicon platform and integrated with conventional CMOS-based circuits.

There are several fundamental drawbacks inherent to the SWB. (i) The speed of signal propagation in the SWB is limited by the spin wave phase velocity, which usually does not exceed $10^5$ m/s [5]. This low phase velocity inevitably results in time delay for signal propagation. The intrinsic SWB delay can be compensated, in part, by decreasing the distances among the inductively coupled circuits. (ii) Another significant drawback is high signal attenuation. Our experimental results for NiFe and CoFe films show that the peak inductive voltage in the output circuit is about 1000 times less than the peak voltage in the input circuit. The signal attenuation is mainly due to the fact that the output voltage is proportional to the spin wave amplitude which carries only a small fraction of the energy from the large excitation voltage. The majority of the input energy is “wasted” in the stray field. In addition, the spin wave signal can only propagate to a finite distance due to damping. The fundamental cause of spin wave amplitude damping is the scattering on conducting electrons, phonons, and magnons. It has been shown experimentally and verified theoretically that magnetic dissipation plays a significant role in propagation of spin waves. Due to the dissipation even in the high quality low-loss yttrium iron garnet (YIG) films the propagation length has been found not to exceed one centimeter. The typical spin wave life-time is limited to the order of a few hundred nanoseconds [10]. This fact restricts the application of the SWB to short-range in-chip interconnects. The coupling efficiency can be improved by optimizing the structure and by using ferrite materials with significantly lower intrinsic attenuation.

## VI. Conclusions

We presented experimental data on inductive coupling via 100 nm thick NiFe and CoFe ferromagnetic films at room temperature. The inductive voltage signal was observed in the output microstrip, as spin wave transit via the ferromagnetic film. Oscillations of the inductive voltage and frequency conversion have been observed. Based on the obtained experimental data, we modeled the logic performance of the inductively coupled circuits. Our numerical simulations show the examples of “NOT” and “AND” logic gates. An original scheme of RTD-based logic cell inductively coupled to the SWB has also been described. The controllable switching is shown by numerical modeling. There are many potential advantages associated with the use of inductively coupled circuits for information processing. The information exchange between the circuits can be accomplished via the spin wave bus, where a bit of information is encoded into the phase or the amplitude of the spin wave signal. The use of spin waves offers an original way to build a new type of micro and nanometer scale logic devices without caring charge current. The issues of low group velocity and damping restrict the potential applications of the SWB to short-range in-chip interconnects.

Acknowledgments

We would like to thank Dr. K. Galatsis for the valuable discussions. The work was supported in part by the MARCO-FENA center, by the Western Institute of Nanoelectronics, and the US Department of Energy.

Figure Captions

Fig.1 (a) Material structure for the prototype logic device with a Magnetic Bus. The core of the structure from the bottom to the top consists of a semiconductor substrate, a ferromagnetic layer, and an insulator layer. There are three microstrips on the top. The edge microstrips act as the input ports, and the one in the middle as the output.

(b) The equivalent circuit schematic for prototype logic device with the Magnetic Bus. There are three inductively coupled oscillators ($L_1,C_1,R_1$), and the transmission line ($L_0,C_0,R_0$) stands for the Magnetic Bus. The phase of oscillation in each circuit depends on the applied voltage and the relative phase of oscillations in two other circuits.

Fig.2 (a) Test structure. The core of the structure from the bottom to the top consists of silicon substrate, 100nm thick ferromagnetic film (NiFe or Cofe), 500 nm silicon oxide layer. There are two microstrip elements on the top of the structure: 2 μm-wide transmission line, 8 μm-wide ground line. The distance between the microstrips is 4 μm. (b) Experimental setup for time-resolved inductive voltage measurements. Picosecond 4000B Pulse Generator was used to bias the excitation antenna. The impedance of the external circuit was matched to 50 Ω. The inductive voltage in the receiving microstrip was detected by a 50 GHz oscilloscope.

Fig.3 (a) Inductive voltage measured for the device with 100nm thick CoFe layer obtained with and without external magnetic field $H_{ext}$=0 and $H_{ext}$=50Oe. (b)The result of the subtraction: signal obtained for $H_{ext}$=0 is subtracted from $H_{ext}$=50Oe.

Fig.4 Inductive voltage measured for the device with 100nm thick NiFe layer obtained with and without external magnetic field $H_{ext}$=0 and $H_{ext}$=50Oe. (b)The result of the subtraction: signal obtained for $H_{ext}$=0 is subtracted from $H_{ext}$=50Oe.

Fig.5 (a) Experimental setup for frequency-domain measurements. Two continuous signals on 6.2 GHz and 6.9 GHz were generated by a Agilent 83711B and a Agilent 83640B signal generators. The set of RF devices were used to mix and amplify the combined signal. Two band pass filters BF1 and BF2 (6.16-7.0 GHz window) are used to cutoff mix frequencies produced by TWT power amplifier. The output signal was detected by the (Agilent 8592A) spectrum analyzer.

Fig.6 Experimental data on frequency-domain measurement. (a) Output signal spectrum in the range of 3-15 GHz. The two most prominent frequency responses (-11 dBm) correspond to the excitation frequencies 6.2 GHz and 6.9 GHz. A signal with a lower power was observed at the frequencies that correspond to the second harmonics of the input signal and the linear combinations of the input signal frequencies and the second harmonics, $2f_2 - f_1$, $2f_1$, $f_1 + f_2$, $2f_2$, *and* $2f_1 - f_2$. (b) Output signal spectrum in the range of 0.6-0.8 GHz. A much lower power signal (-62 dBm) was detected at the frequency $f_2 - f_1$.

Fig.7 (a) An example of digital circuits connected via the spin wave bus. With a constant bias $V_{DD}$, each circuit may have two possible stable states: “low” and “high” output voltages, corresponding to the logic states 0 and 1, respectively. (b) Approximate tunneling characteristics of a tunneling diode. There are three linear portions of the approximated curve, labeled as *b1*, *b2*, and *b3*. $V_p$, $I_p$, $V_v$, $I_v$ are the peak voltage, the peak current, the valley voltage, and the valley current, respectively. The slopes of the branches *b1* and *b3* are the same.

Fig.8 Results of numerical simulations showing selective switching of the bi-stable RTD-based circuit by the inductive voltage signal. (a) Inductive voltage signal (shown in the inset) does not switch the state of the RTD-based circuit. (b) Time dependent bias $V_{DD}$ is synchronized with the inductive voltage signal. The circuit switches its state from the “low” output voltage to the “high” output voltage by the inductive signal. Voltage and time are normalized to $V_p$ and RC, respectively, where C is the circuit capacitance and $R=V_p/I_p$, $I_p$ the RTD peak current.

Fig.9 Logic gates implementation in the inductively coupled circuits. (a) “NOT” gate as a delay line. The signal transmission time from the input to the receiver can be adjusted to provide a $\pi$ phase difference between the input and output voltage signals. (b) ”AND” gate as a result of the superposition of two signals. The magnitude of the output signal is a function of the input signals relative phase. (c) “AND” gate using the frequency conversion. The output signal at frequency $f_1+f_2$ is nonzero if and only if there are nonzero inputs at both input frequencies $f_1$ and $f_2$ .

References

1. ITRS, Semiconductor Industry Association, 2005. **http://www.itrs.net/Common/2005ITRS/Home2005.htm**.
2. Davis, J.A., R. Venkatesan, A. Kaloyeros, M. Beylansky, S.J. Souri, K. Banerjee, K.C. Saraswat, A. Rahman, R. Reif, and J.D. Meindl, *Interconnect limits on gigascale integration (GSI) in the 21st century.* Proceedings of the IEEE, 2001. **89**(3): p. 305-24.
3. Schilz, W., *Spin-wave propagation in epitaxial YIG films.* Philips Research Reports, 1973. **28**(1): p. 50-65.
4. Silva, T.J., C.S. Lee, T.M. Crawford, and C.T. Rogers, *Inductive measurement of ultrafast magnetization dynamics in thin-film Permalloy.* Journal of Applied Physics, 1999. **85**(11): p. 7849-62.
5. Covington, M., T.M. Crawford, and G.J. Parker, *Time-resolved measurement of propagating spin waves in ferromagnetic thin films.* Physical Review Letters, 2002. **89**(23): p. 237202-1-4.
6. Khitun, A. and K. Wang, *Nano scale computational architectures with Spin Wave Bus.* Superlattices & Microstructures, 2005. **38**: p. 184-200.
7. Kostylev, M.P., A.A. Serga, T. Schneider, B. Leven, and B. Hillebrands, *Spin-wave logical gates.* Applied Physics Letters, 2005. **87**(15): p. 153501-1-3.
8. Allwood, D.A., G. Xiong, C.C. Faullkner, D. Atkinson, D. Petit, and R.P. Cowburn, *Magnetic domain-wall logic.* Science, 2005. **309**(5741): p. 1688-92.
9. Chua, L.O. and L. Yang, *Cellular neural networks: theory.* IEEE Transactions on Circuits & Systems, 1988. **35**(10): p. 1257-72.
10. Kalinikos, B.A. and M.P. Kostylev. *Parametric amplification of spin wave envelope solitons in ferromagnetic films by parallel pumping*. in *IEEE. IEEE Transactions on Magnetics, vol.33, no.5, pt.1, Sept. 1997, pp.3445-7. USA.*

a)

b)

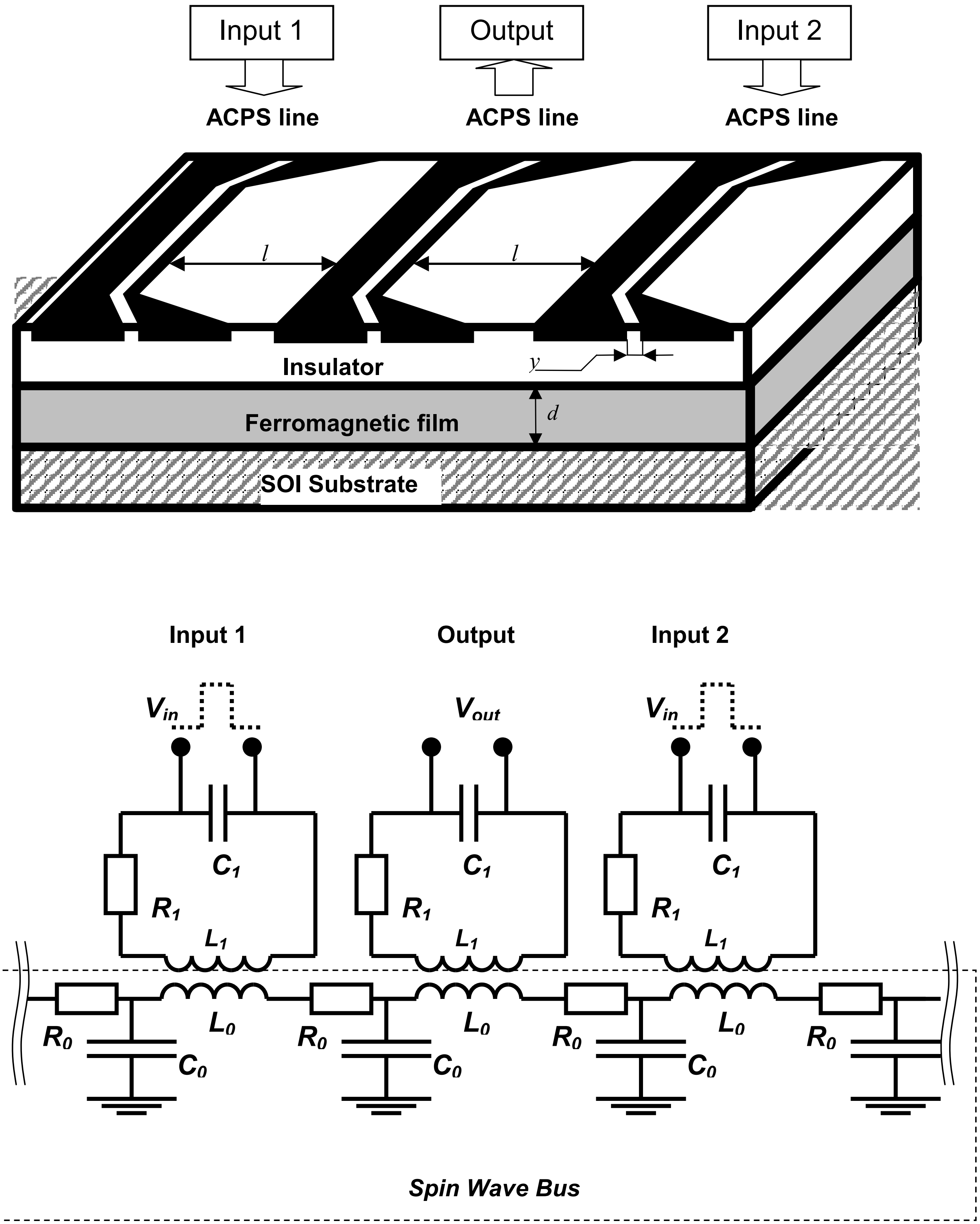

Fig.1

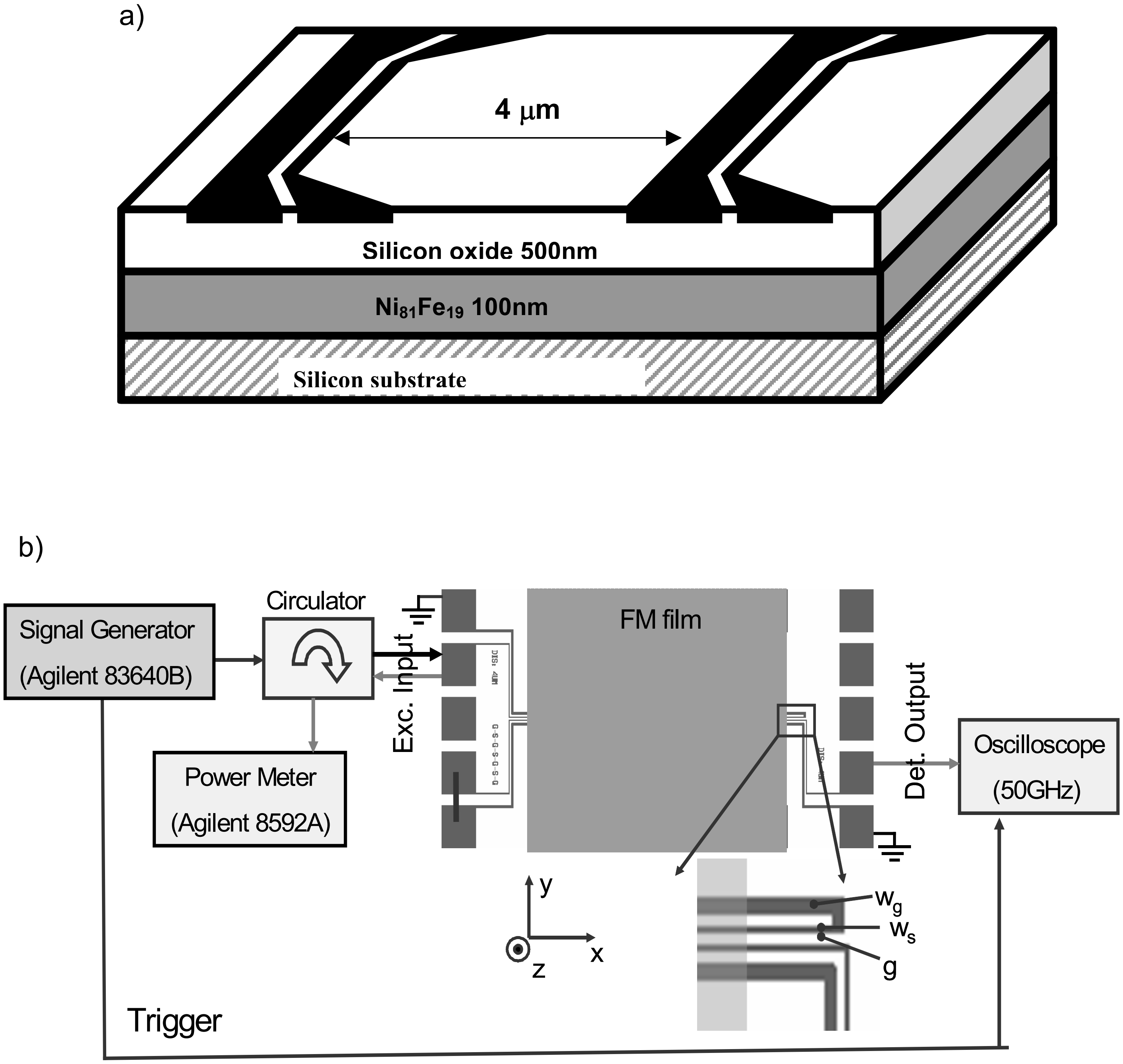

Fig.2

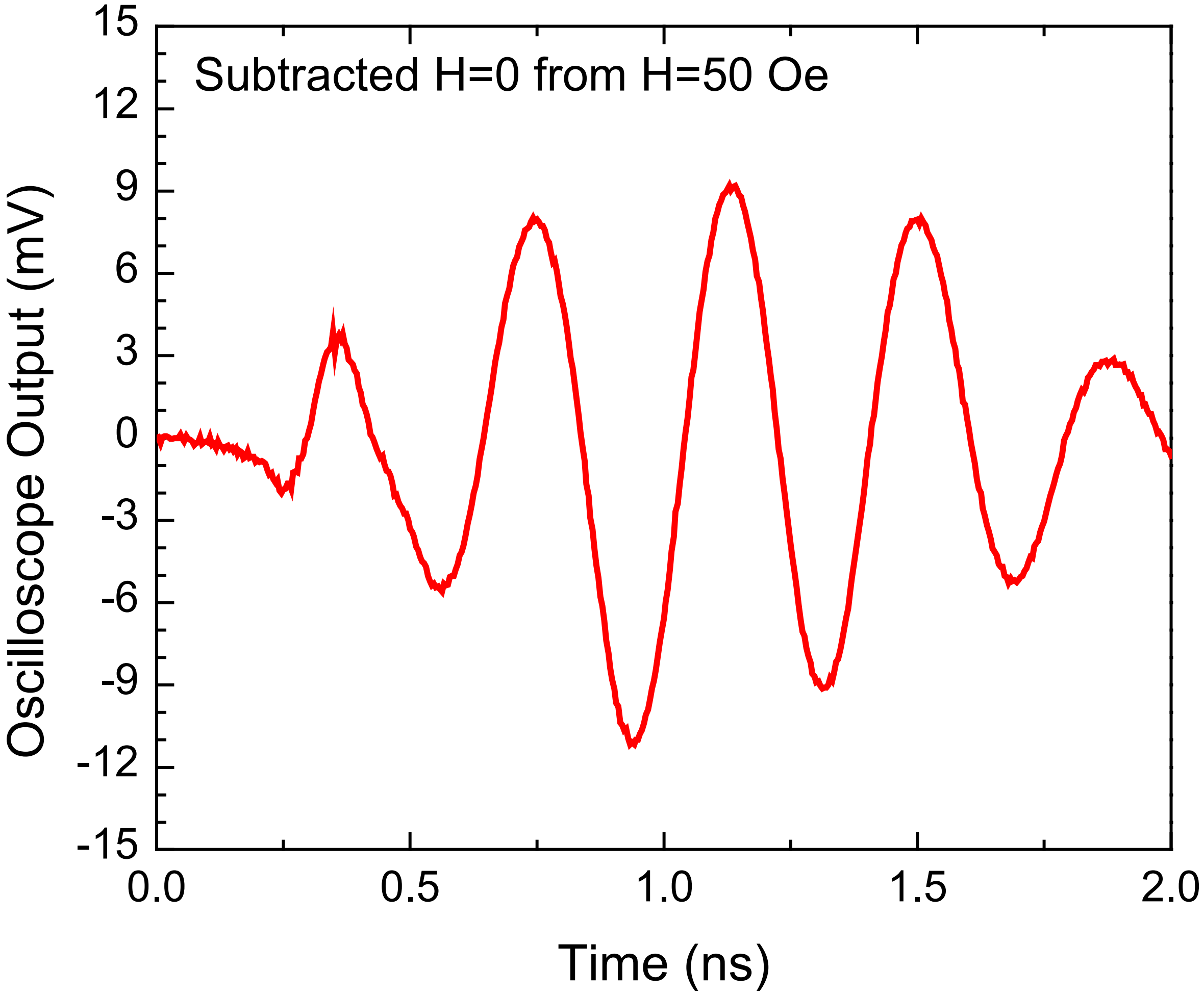

Fig.3

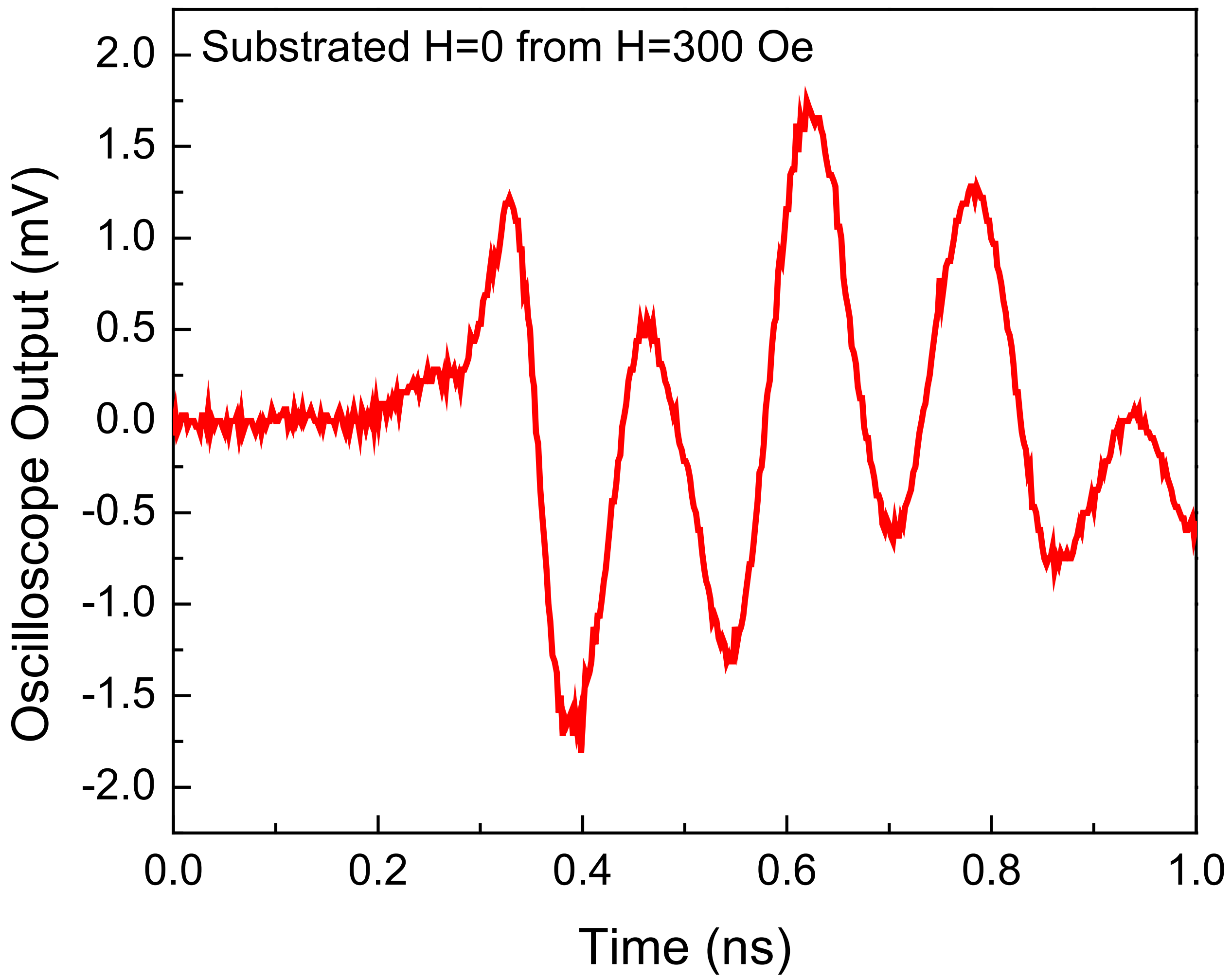

Fig.4

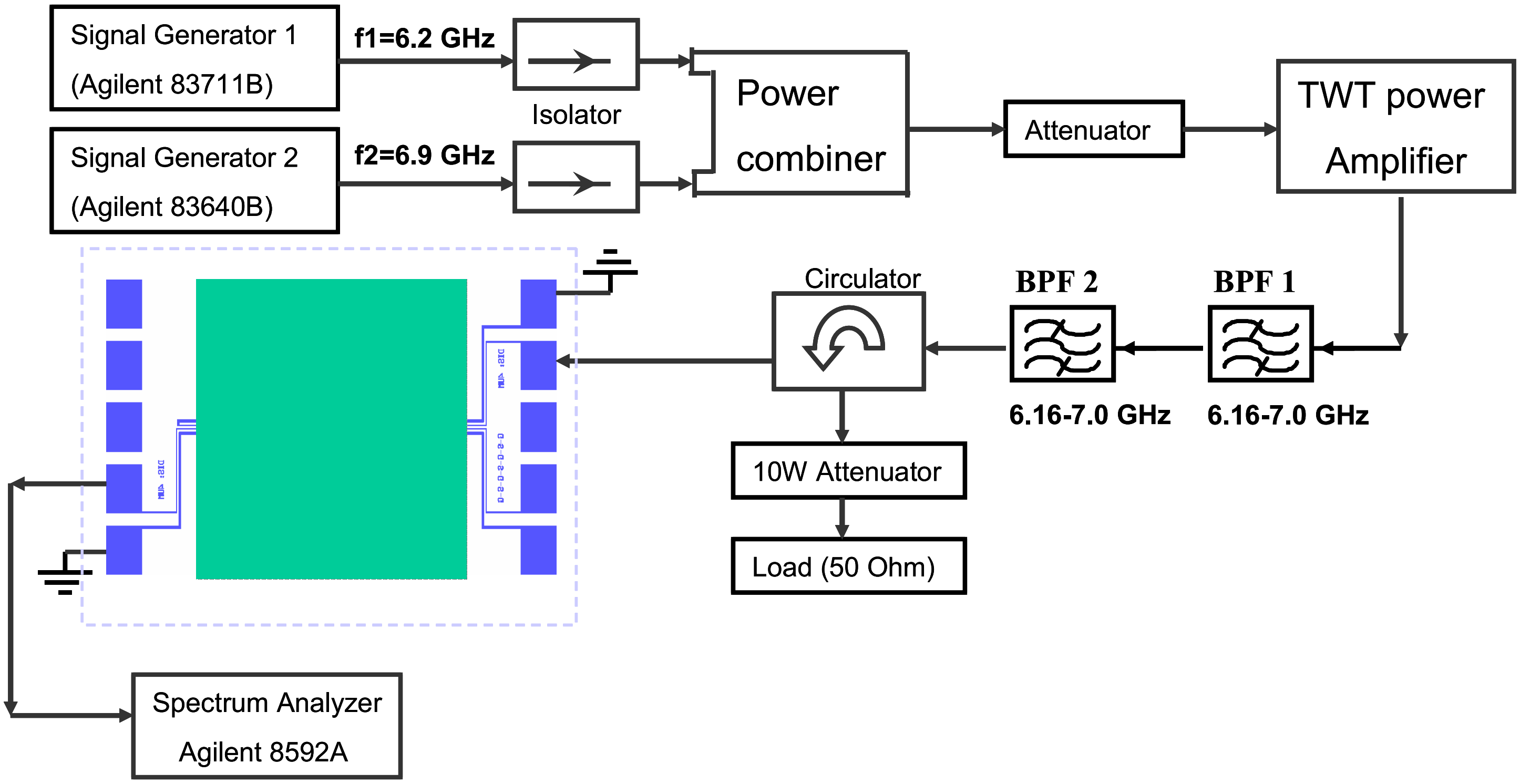

Fig.5

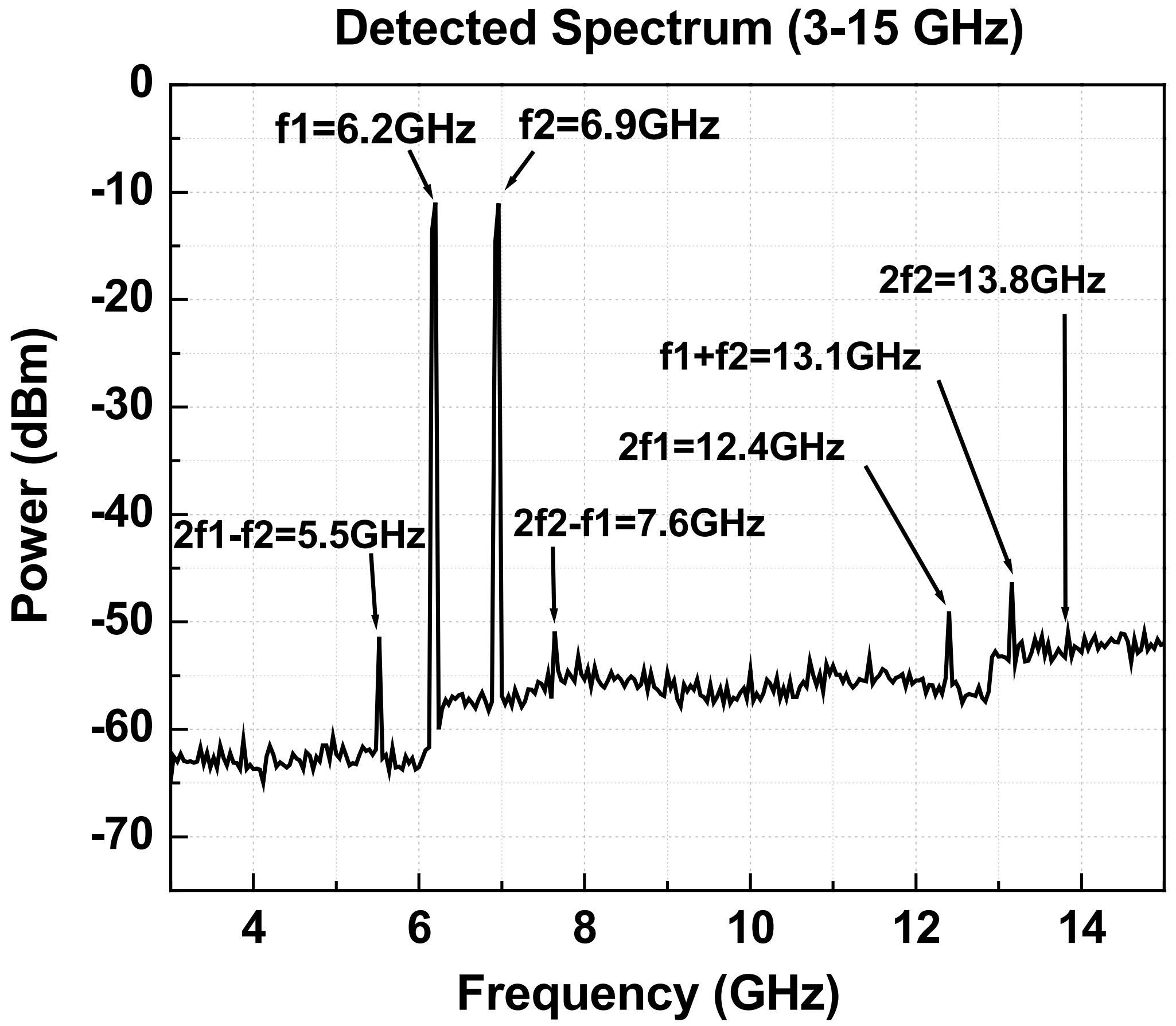

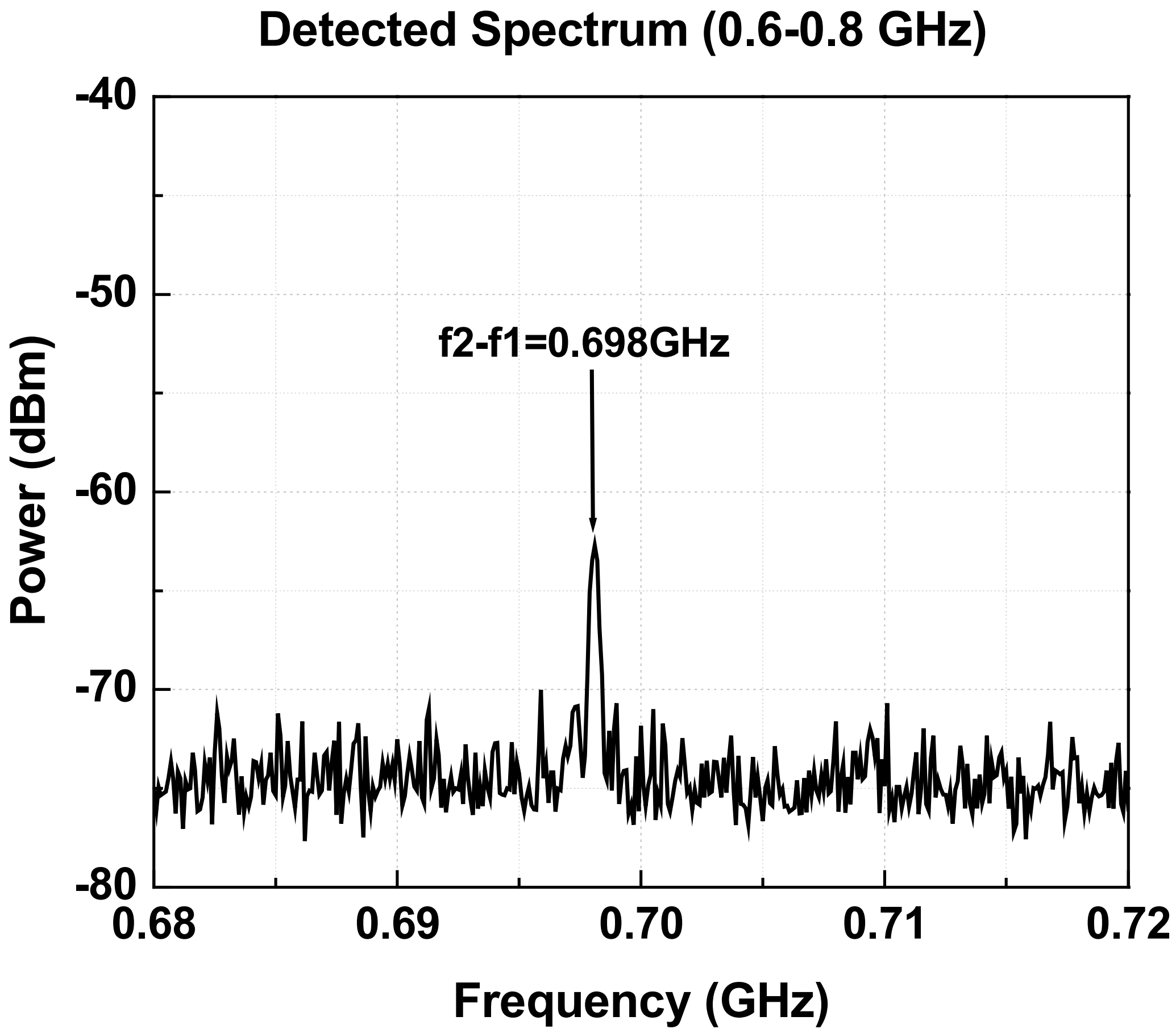

Fig.6

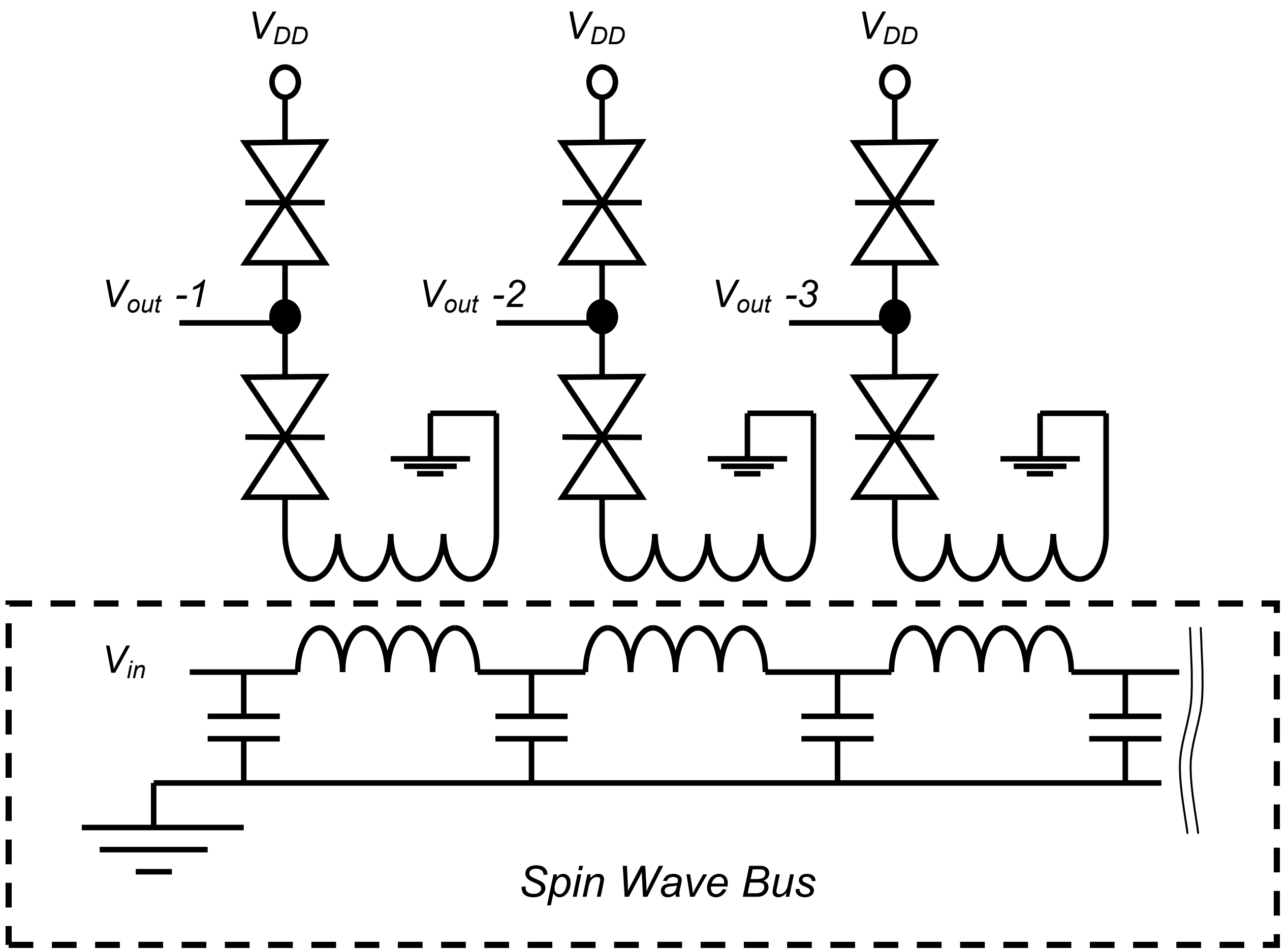

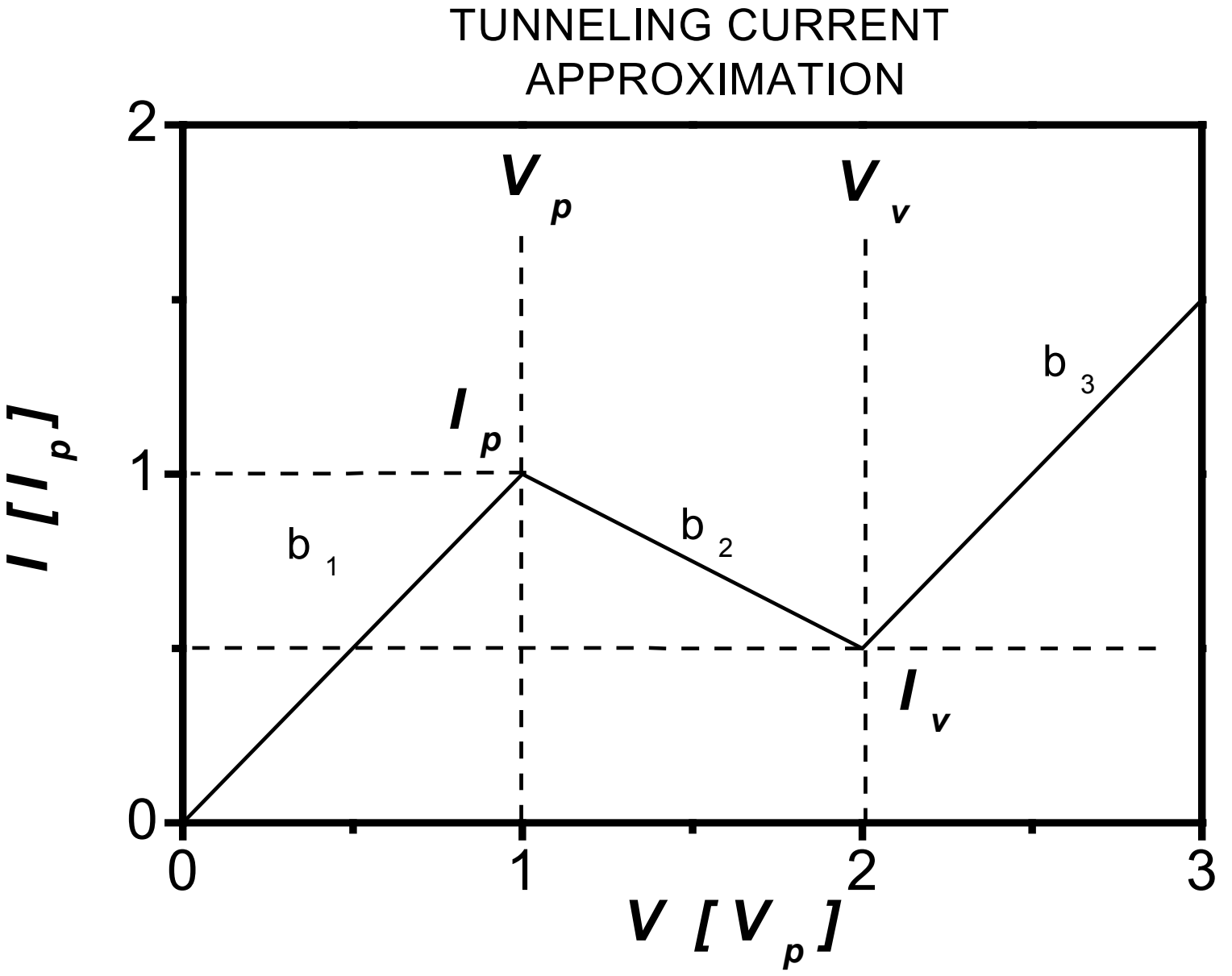

Fig.7

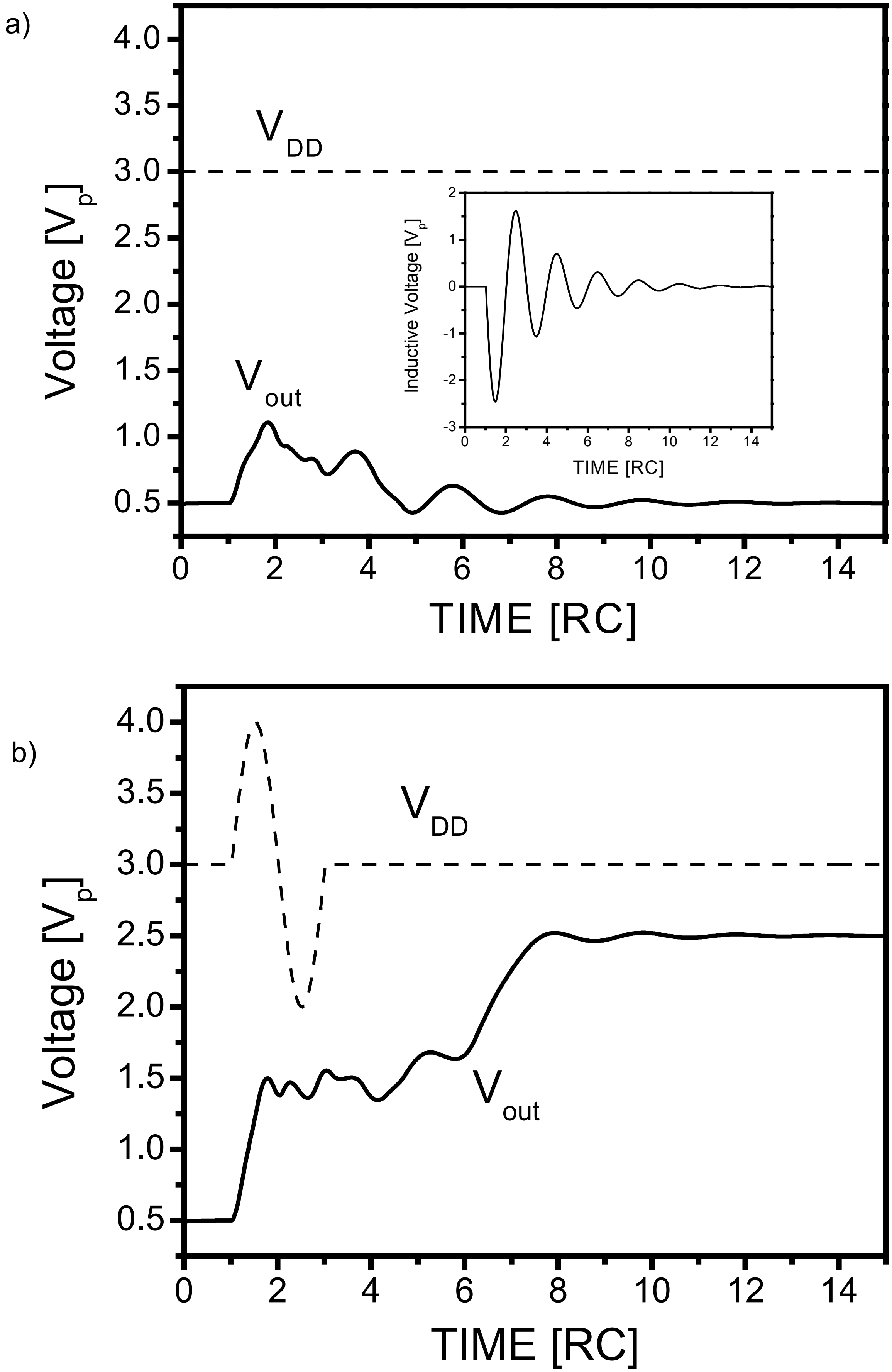
a)
Voltage [Vp]
4.0
3.5
3.0
2.5
2.0
1.5
1.0
0.5
VDD
Vout
Inductive Voltage [Vp]
2
1
0
-1
-2
-3
0 2 4 6 8 10 12 14
TIME [RC]
0 2 4 6 8 10 12 14
TIME [RC]
b)
Voltage [Vp]
4.0
3.5
3.0
2.5
2.0
1.5
1.0
0.5
VDD
Vout
0 2 4 6 8 10 12 14
TIME [RC]

Fig.8

a)

**NOT gate (delay line)**

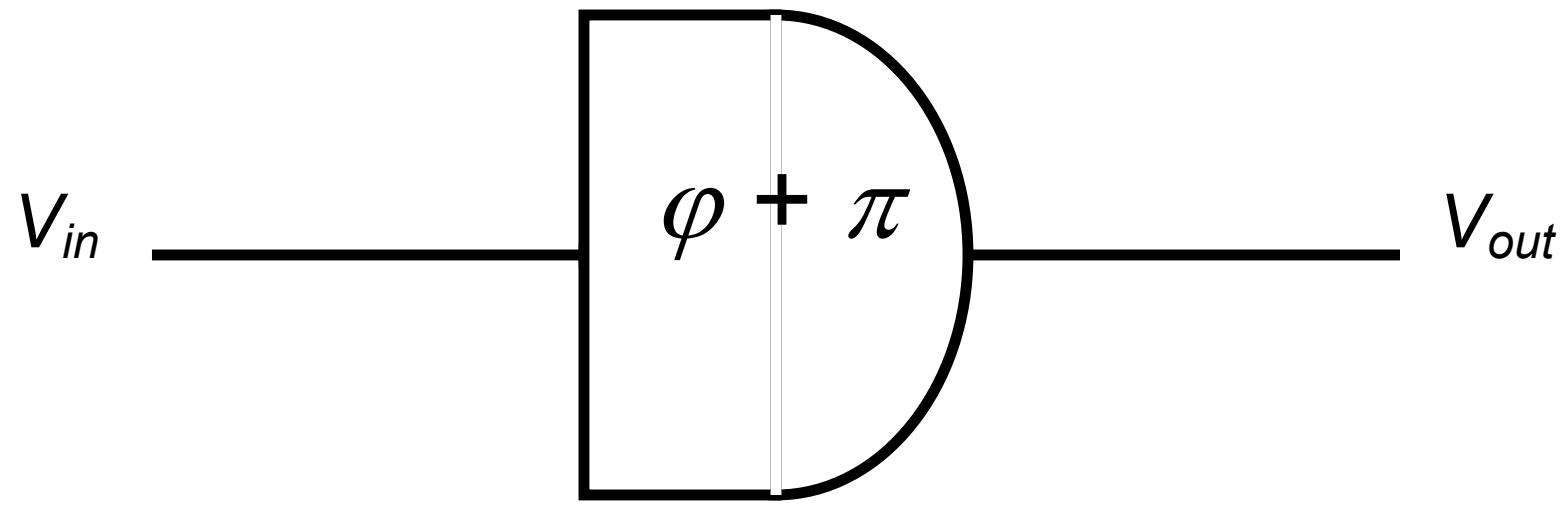

$$V_{out} = V_{in} \cdot \sin(\omega t_{prop})$$
$$\omega t_{prop} = \pi$$

| Input | Output |
|---|---|
| 0 | 1 |
| 1 | 0 |

b)

**AND gate (wave superposition)**

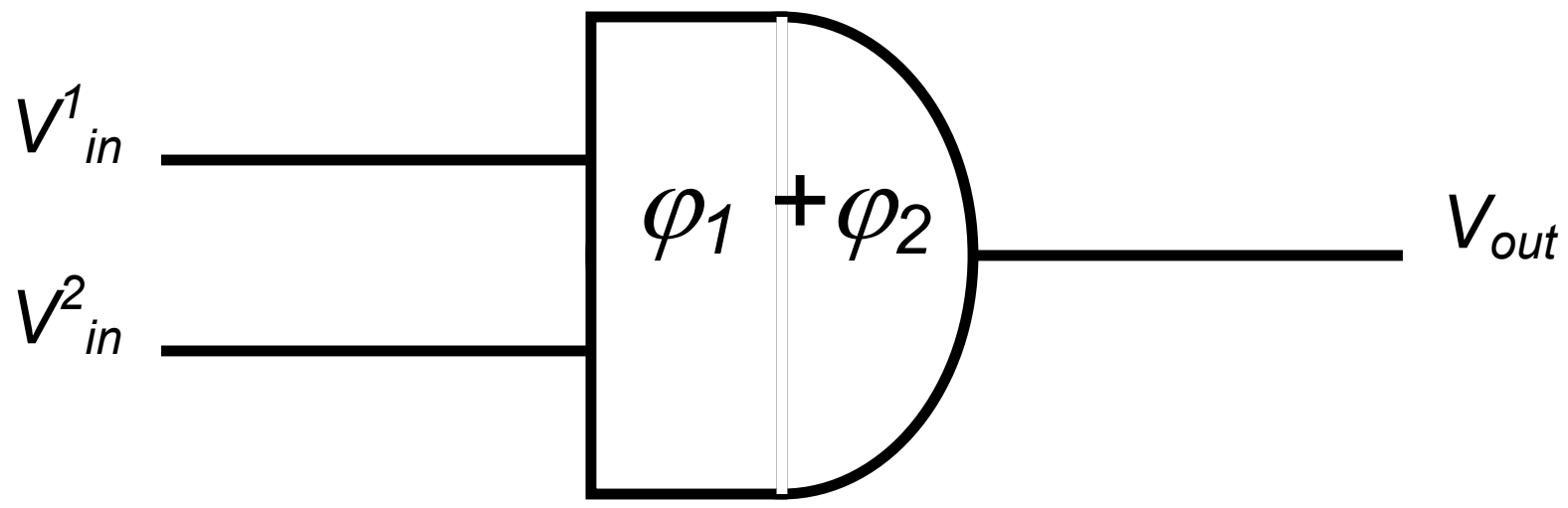

$$V_{out} = \sqrt{V_1^{\,2} + V_2^{\,2} + 2V_1 V_2 \cos(\varphi_2 - \varphi_1)}$$

| Input | Output |
|---|---|
| 1 0 | 0 |
| 1 0 | 0 |
| 0 1 | 0 |
| 1 1 | 1 |

c)

**AND gate (frequency conversion)**

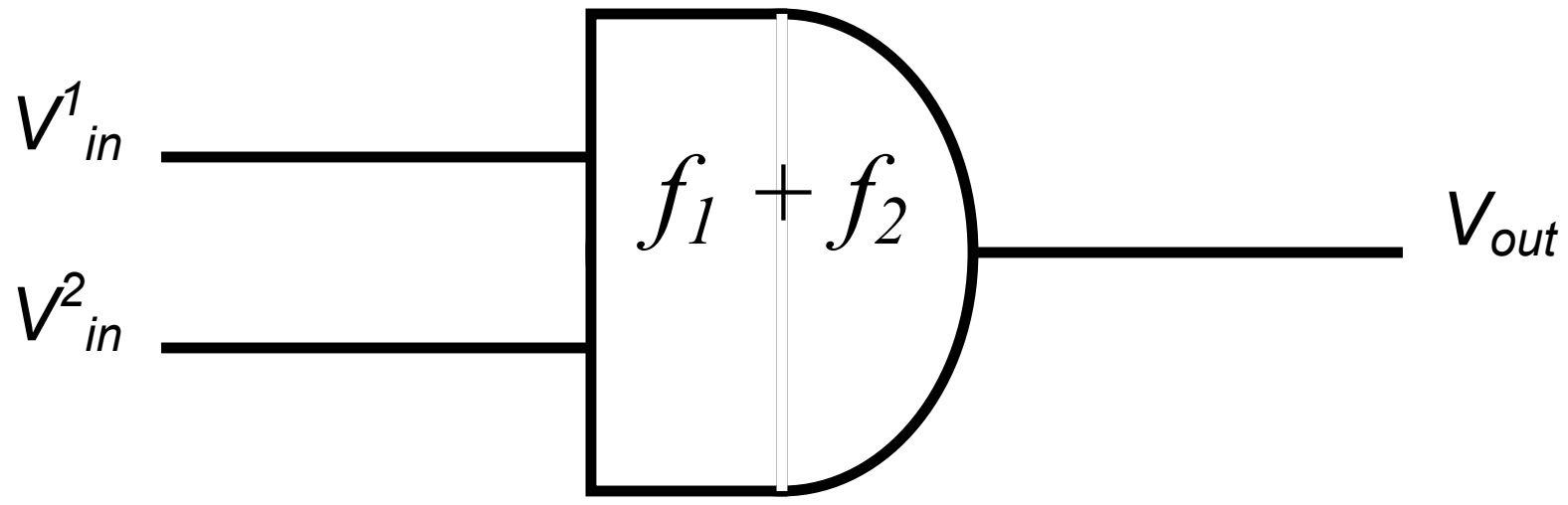

$$V_{out}(f_1 + f_2) = \eta \cdot V_1(f_1) \cdot V_2(f_2)$$

| Input | Output |
|---|---|
| 0 0 | 0 |
| 1 0 | 0 |
| 0 1 | 0 |
| 1 1 | 1 |

Fig.9